# A Review of Modeling and Waveform Inversion for Marine Seismic Data

**Guoxin Chen**
Ocean College, Zhejiang University, 316021, Zhoushan, China

## Abstract

Marine seismic exploration serves as a core technology supporting marine resource exploration, seabed structural detection, carbon sequestration monitoring, and offshore engineering safety. The deep integration of full-waveform inversion (FWI), elastic inversion, numerical modeling, and artificial intelligence is driving the paradigm shift of marine seismic exploration from traditional physics-driven methods to a **physics-constrained and data-driven dual-drive mode**. Based on the special issue *Modeling and Waveform Inversion of Marine Seismic Data* in the **Journal of Marine Science and Engineering (JMSE)**, this paper systematically reviews 11 high-quality papers collected in this issue. Six technical branches are covered in detail: seismic data preprocessing, forward modeling, FWI, elastic inversion, reservoir characterization, and migration imaging. Each paper is elaborated in terms of scientific problems, methodological innovations, core contributions, and application values, presenting a complete technical chain, breakthrough paths, and development trends in marine seismic modeling and waveform inversion.

The results show that intelligent reconstruction of irregular sampling, multi-source data joint inversion, low-frequency missing and cycle-skipping suppression, physics-guided deep-learning inversion, and wide-spectrum velocity modeling have become key technical routes to overcome industry bottlenecks, targeting complex observation scenarios such as ocean-bottom nodes (OBN/OBC), streamers, and passive sources. The achievements of this special issue form a complete system from theoretical innovation and methodological breakthroughs to engineering verification, providing solid theoretical and technical support for national strategic demands including deep-water oil and gas exploration, seabed hazard detection, and deep-sea carbon sequestration dynamic monitoring. Finally, this paper introduces the theme and scope of the new JMSE special issue *Marine Geophysical Exploration in the Era of Artificial Intelligence*, incorporating the author's recent work on AI applications in seismic exploration, and prospects the future direction of the deep integration of AI and marine seismic exploration.

**Keywords**: marine seismic data; forward modeling; full-waveform inversion; elastic inversion; deep learning; multi-data fusion; OBN/OBC; passive-source seismology

---

## 1 Introduction

Global marine resource development and deep-sea scientific research have entered a stage of high-precision, intelligent, and quantitative development. The exploration of seabed oil and gas resources, natural gas hydrates, and mineral resources, as well as crustal structure detection, geological disaster early warning, and carbon capture and storage (CCS) monitoring, all impose higher requirements on marine seismic imaging and inversion. Marine seismic data are constrained by numerous natural factors during acquisition: sparse spatial sampling of ocean-bottom observation systems (OBN/OBC), limited coverage of streamer data, irregular sampling of passive-source data, severe lack of low-frequency signals, strong scattering from seawater and sediment layers, and cycle-skipping and local minima caused by complex structures. As a result, traditional seismic processing and inversion methods struggle to obtain high-fidelity, high-resolution subsurface velocity and elastic parameter models.

Full-waveform inversion (FWI) achieves high-precision inversion of subsurface medium parameters by matching observed and synthetic wavefields, making it one of the most promising techniques in seismic exploration. Elastic wave inversion can simultaneously obtain key reservoir parameters including P-wave velocity, S-wave velocity, density, Poisson's ratio, and Young's modulus. Numerical forward modeling provides a physical basis for wavefield propagation analysis, error correction, and inversion objective function construction. Intelligent methods represented by deep learning and Bayesian estimation offer efficient tools for data reconstruction, noise suppression, inversion initialization, and multi-source information fusion.

The special issue *Modeling and Waveform Inversion of Marine Seismic Data* in JMSE is guest-edited by Guoxin Chen, Chunfeng Li from Zhejiang University, and Yangting Liu from the First Institute of Oceanography, MNR. As of May 2026, it includes 11 academic papers with 19 total citations, 16,888 total views and downloads, over 1,500 average views per paper, and 1.7 average citations per paper. The issue covers the **full workflow of forward modeling, data preprocessing, core inversion, intelligent empowerment, and engineering applications** in marine seismology. It focuses on three mainstream exploration scenarios: OBN/OBC, streamer, and passive sources, and addresses common industrial challenges such as low-frequency absence, cycle-skipping, irregular sampling, and multi-data fusion, representing the frontier of international marine seismic waveform inversion. This paper reviews each of the 11 papers in detail, extracts scientific problems, technical innovations, and application values, forms a complete research progress summary, presents recent advances in AI applications in marine geophysics from the author's group, and prospects future trends combined with the new upcoming special issue.

---

# 2 Intelligent Preprocessing and Reconstruction of Marine Seismic Data

## 2.1 Intelligent Interpolation of OBN Multi-Component Seismic Data Using a Frequency-Domain Residual-Attention U-Net

Ocean-bottom node (OBN) acquisition systems use multi-component seismic data to record richer wavefield information than conventional streamer data, including P-waves, S-waves, and converted waves, showing significant advantages in imaging deep-water complex structures. However, limited by seabed topography, exploration costs, and deployment conditions, OBN multi-component data commonly suffer from **sparse spatial sampling, irregular distribution, and severe data missing**, which directly lead to spatial aliasing, reduced resolution in seismic imaging, and local minima in waveform inversion due to discontinuous data, failing to meet high-precision exploration demands. Traditional interpolation methods such as Kriging, polynomial interpolation, and f-x inversion rely on linear assumptions and lack accuracy in non-stationary, highly scattered, low signal-to-noise ratio marine seismic data, especially for data reconstruction near complex geological boundaries and faults.

To solve these problems, **Zhang and Yu (2026)** proposed a U-Net deep learning model integrating **frequency-domain features** and **residual-attention mechanism** for intelligent interpolation and signal enhancement of OBN multi-component seismic data. The method transforms seismic data from the time–space domain to the frequency domain to better characterize the periodicity and coherence of seismic signals and reduce nonlinear interference. Residual connections are introduced into U-Net to alleviate gradient vanishing during deep network training and improve convergence speed and stability. Meanwhile, an attention module is embedded to automatically focus on effective signal regions, suppress noise and redundant information, and enhance learning and reconstruction of key geological features such as faults and boundaries.

Experimental results demonstrate that the method achieves high-precision reconstruction of irregularly and sparsely sampled OBN data, significantly improving data continuity, signal-to-noise ratio, and coherence. The reconstructed results can be directly used for subsequent migration imaging and full-waveform inversion. Compared with traditional interpolation methods, the model maintains high robustness under complex seabed structures and low SNR conditions, providing a standardized preprocessing scheme for OBN data and important engineering value for improving the accuracy of deep-water OBN exploration.

## 2.2 Full Waveform Inversion of Irregularly Sampled Passive Seismic Data Based on Robust Multi-Dimensional Deconvolution

Passive-source marine seismic exploration uses ambient noise, natural earthquakes, and other passive signals to image subsurface structures without artificial sources. It features low cost, low environmental impact, and sustainable observation, showing broad application prospects in deep-sea long-term monitoring, carbon sequestration assessment, and geological disaster early warning. However, passive-source data face challenges including **unknown source locations, weak signal energy, irregular sampling, and strong noise interference**. Conventional passive-source inversion methods struggle to achieve stable velocity model reconstruction; sparse and non-uniform data distribution directly leads to low inversion accuracy and convergence difficulties.

To address these issues, **Zhang et al. (2025)** proposed a **Robust Multi-Dimensional Deconvolution (RMDD)** algorithm and an adaptive passive-source full-waveform inversion strategy. The method first reconstructs passive-source signals with high precision via RMDD to suppress environmental noise and coherent interference. Then, a multi-domain interpolation strategy (time–space domain + frequency domain) is adopted to fill missing data in irregular sampling and improve data uniformity and integrity. On this basis, an adaptive FWI workflow is designed to dynamically adjust regularization weights and iteration strategies according to data quality, reducing dependence on initial models and improving inversion stability.

The core contribution of this method is that multi-domain joint interpolation significantly improves virtual source reconstruction accuracy, enabling stable and high-precision inversion of passive-source data in complex marine environments. Compared with conventional active-source exploration, it greatly reduces costs and environmental disturbance, suitable for special scenarios such as deep water, open sea, and ecologically sensitive areas. This research provides key technical support for the practical application of passive-source marine seismic data and broadens the technical boundary of marine seismic observation.

### 2.3 Improving Seismic Impedance Inversion by Fully Convolutional Neural Network

Seismic impedance inversion is a core link connecting seismic data to reservoir lithology and physical properties, directly determining the accuracy of reservoir prediction and fluid identification. Deep-learning-based impedance inversion offers advantages of high speed and strong adaptability, but existing models commonly suffer from **weak generalization, poor adaptability to field data, insufficient recovery of high-frequency information, and difficulty preserving low-frequency trends**, failing to maintain stable and high precision under different work areas and signal-to-noise ratios, which limits industrial application.

**Tao et al. (2025)** proposed a joint inversion strategy of **histogram equalization + fully convolutional U-Net** to improve impedance inversion accuracy and generalization from both data enhancement and network structure optimization. First, histogram equalization is used to normalize amplitudes and enhance features of seismic data, strengthening effective reflection characteristics, weakening amplitude distortion caused by acquisition and environment, and improving data consistency. Then, a fully convolutional U-Net is constructed to achieve end-to-end mapping from seismic signals to impedance, using an encoder to extract multi-scale deep features and a decoder to output high-resolution results while preserving detailed information.

The most significant innovation is that **broadband inversion of field marine seismic data can be realized using only synthetic data for training**, avoiding limitations of insufficient labeled field data and scarce well-log information, greatly improving practicality. Experimental results show that the method effectively broadens inversion bandwidth, recovers low-frequency background trends and high-frequency details, and the inversion results are highly consistent with well-log data, providing a low-cost, high-generalization solution for intelligent seismic impedance inversion in marine exploration.

---

## 3 Forward Modeling and Wavefield Propagation Analysis

### 3.1 Numerical Modeling on OBS P-Wave Receiver Function to Analyze Influences of Seawater and Sedimentary Layers

The Ocean-Bottom Seismograph (OBS) receiver function method is an important tool for studying deep crustal structures, Moho depth, and lithospheric properties, widely used in oceanic crust detection. However, unlike continental environments, **seawater and surficial sediment layers** generate strong scattering and multiple-wave interference on teleseismic P-waves, distorting receiver function waveforms and severely reducing the accuracy of deep structural imaging. Previous studies have not systematically quantified the impacts of seawater and sediment thickness on receiver functions, lacking reliable correction basis and leading to systematic biases in inversion results.

**Gong et al. (2024)** constructed an accurate marine geophysical theoretical model and conducted systematic numerical parametric tests, changing key parameters such as seawater thickness, sediment thickness, and sediment velocity to quantitatively analyze their influences on OBS P-wave receiver function waveforms, amplitudes, and arrival times. The results clearly show that reverberations from the seawater layer have a **negligible effect** on receiver functions, while the sediment layer response is the **main cause** of Moho depth estimation bias. Thicker sediments and larger velocity contrasts lead to more severe distortion and greater errors in deep interface identification.

This study provides key theoretical basis and quantitative correction methods for OBS receiver function data calibration, effectively improving the accuracy of marine deep crustal structure inversion and having important scientific significance for global oceanic tectonic research.

## 3.2 Hierarchical Joint Elastic Full Waveform Inversion Based on Wavefield Separation

Single observation systems (streamer or OBN) have inherent defects: streamer data mainly record up-going wavefields and lack S-wave information; OBN data are multi-component but have limited coverage density. Independent inversion using either type struggles to build high-precision subsurface velocity models, especially in complex structural areas where multi-solution and local minima frequently occur. Meanwhile, conventional elastic FWI faces problems such as **strong P–S wave coupling, unstable gradient calculation, and severe acquisition footprint interference(Chen et al., 2022)**.

**Han et al. (2025a)** proposed a **hierarchical joint elastic full-waveform inversion method (PS-JFWI)** based on wavefield separation to complement advantages of streamer and OBN data. The method first separates P and S waves from multi-component data to reduce coupling interference. Then, an **acoustic–elastic coupled wave equation** and **gradient decoupling algorithm** are constructed to achieve accurate simultaneous inversion of P- and S-wave velocities. A hierarchical inversion strategy is adopted, optimizing the model from shallow to deep and from background to details, alleviating cycle-skipping and local minima.

Experiments show that the method effectively eliminates acquisition footprints, improves velocity model accuracy in complex structural areas, and solves the problem of insufficient information in single-data inversion, providing a standardized physical framework and algorithm workflow for multi-source data joint inversion.

---

# 4 Innovative Methods for FWI and Elastic Wave Phase Inversion

## 4.1 Elastic Wave Phase Inversion in Local-Scale Frequency–

## Wavenumber Domain

Marine towed simultaneous-source exploration features high efficiency and low cost but suffers from **serious low-frequency data missing** (Chen et al., 2018, Luo et al., 2019). Conventional FWI relies on low frequencies to build background velocity models; low-frequency absence directly causes **cycle-skipping**, trapping inversion in local minima and failing to recover correct long-wavelength background models. In addition, amplitude information is sensitive to noise, source wavelets, and coupling effects, leading to poor stability of amplitude-matching FWI.

**Qu et al. (2025)** proposed a **local-scale frequency–wavenumber domain elastic wave phase inversion (LFKEPI)** method. The core idea is to **abandon amplitude matching and focus on phase matching**, constructing a phase misfit function insensitive to amplitude noise to extract long-wavelength background velocity information from phase data. The method performs local phase calculation in the frequency–wavenumber domain, reducing global search cost and improving inversion efficiency and stability, especially suitable for marine towed simultaneous-source data.

This method breaks the strong dependence of conventional FWI on low-frequency data, enabling high-quality initial model construction even without low frequencies, laying a foundation for subsequent high-precision FWI and representing a landmark innovation for solving cycle-skipping.

## 4.2 Physics-Guided Self-Supervised Learning FWI with Simultaneous Source Pretraining

Conventional FWI highly depends on initial model accuracy and easily fails under low-frequency missing and complex structures. Pure data-driven deep-learning inversion lacks physical constraints, producing results inconsistent with the wave equation and showing poor generalization and weak interpretability. How to organically integrate **physical mechanisms** and **deep learning** to build a stable, efficient, and noise-robust FWI framework is a core challenge in intelligent marine seismic inversion.

**Zheng et al. (2025)** proposed a two-stage **physics-guided self-supervised learning FWI**: in the first stage, large-scale pretraining is performed using simultaneous-source data to learn general geological features and wavefield laws; in the second stage, fine-tuning is conducted using single-source data to adapt to specific geological conditions of the work area. For network structure, U-Net is improved with a **Partial Convolution Attention Module (PCAM)** to achieve accurate feature extraction under sparse and irregular sampling. Meanwhile, the wave equation is embedded into the loss function as a constraint to ensure physical consistency of inversion results.

Results show that the method effectively avoids cycle-skipping, exhibits excellent robustness to low-frequency absence and strong noise, and significantly outperforms conventional FWI and pure data-driven models in inversion accuracy, convergence speed, and stability, providing an engineering-feasible technical route for intelligent FWI.

## 4.3 Wave Equation Reflection Inversion for Wide-Spectrum Velocity Model Reconstruction

Limited by observation systems, conventional marine seismic surveys face difficulties in acquiring **long-offset data and severe low-frequency component missing**, leading to low resolution of deep velocity models and blurred structural imaging, failing to meet deep-water and deep-buried exploration demands. Conventional reflected-wave inversion methods have narrow bandwidth and high sensitivity to noise, making wide-spectrum velocity reconstruction difficult.

**Ni et al. (2025)** proposed a **Wave Equation Reflection Inversion (WERI)** framework to reconstruct wide-spectrum velocity models using reflected-wave information, reducing strong dependence on long-offset and low-frequency data. Based on the wave equation, the method builds a reverse-time imaging and inversion workflow for reflected waves, integrating travel-time and waveform constraints to achieve high-precision velocity model reconstruction from shallow to deep layers while improving vertical and horizontal resolution.

Verified on the Sigsbee2b benchmark model and East China Sea field data, the WERI method significantly improves deep velocity model accuracy and structural imaging quality, with important application value for deep-water oil–gas exploration and crustal structure detection.

### 4.4 Integrating Multimodal Deep Learning with Multipoint Statistics for 3D Crustal Modeling

Single observation systems (streamer/OBN) cannot comprehensively obtain multi-parameter information such as P-wave velocity, S-wave velocity, and density of complex seabed formations, leading to problems of **incomplete information, difficulty in multi-scale feature fusion, and high uncertainty** in 3D crustal modeling. Traditional geostatistical methods struggle to handle nonlinearly correlated parameters, while deep-learning methods lack geological constraints, producing models inconsistent with real geological laws.

**Liu et al. (2024)** constructed a **joint multi-parameter FWI framework in acoustic–elastic media (J-AEFWI)** to integrate advantages of streamer and OBN data. Multimodal deep learning is combined with multipoint geostatistics, and inversion weights and iteration strategies are optimized to achieve high-precision 3D crustal structure modeling. The method fully utilizes physical laws of seismic data and spatial structure of geostatistics, improving model rationality and accuracy.

Verified on field data from the South China Sea, the method significantly improves multi-parameter inversion accuracy and characterizes the spatial distribution of complex seabed crustal structures, providing reliable 3D models for deep-water oil–gas exploration and regional geological research.

---

## 5 Reservoir Characterization and Migration Imaging

### 5.1 Reservoir Characterization Based on Bayesian AVO Inversion of Marine Seismic Data

Amplitude Versus Offset (AVO) inversion is a core technique for reservoir fluid identification,

lithology classification, and brittleness evaluation. Conventional AVO inversion usually assumes a **constant P-wave to S-wave velocity ratio (Vp/Vs)**, which deviates from actual complex seabed geological conditions, causing systematic errors and difficulty in accurately estimating key parameters such as density, Poisson's ratio, and Young's modulus, limiting the ability to characterize reservoir heterogeneity.

**Wang et al. (2025)** reconstructed a **four-term PP-wave reflectivity equation with density** to replace the traditional three-term approximation, improving reflectivity calculation accuracy. A Cauchy prior distribution was adopted to better characterize non-Gaussian distributions of geological parameters, and a robust Bayesian inversion framework was constructed combined with low-frequency regularization constraints to achieve simultaneous estimation of multiple elastic parameters.

The method accurately estimates key reservoir parameters including Young's modulus, Poisson's ratio, and density, effectively characterizing the heterogeneity of complex marine reservoirs, providing high-precision parameter support for oil–gas reservoir evaluation, hydraulic fracturing design, and carbon sequestration reservoir assessment.

## 5.2 Imaging OBS Data with Acoustic Kirchhoff Pre-Stack Depth Migration

OBS multi-component data feature complex wavefield propagation paths, developed multiples, and uneven observation geometry. Conventional time migration and simple depth migration methods suffer from **low imaging accuracy, severe crosstalk artifacts, and distorted imaging of complex structures**, failing to meet oil–gas exploration and hazard assessment demands. Achieving high-precision OBS imaging and suppressing crosstalk and artifacts is a key challenge in marine seismic imaging.

**Han et al. (2025b)** adopted **Acoustic Kirchhoff Pre-Stack Depth Migration (PSDM)**. Based on ray theory, travel times are traced to adapt to complex subsurface geometry and velocity variations, realizing high-precision depth imaging of OBS data. Numerical simulations systematically analyze migration responses and crosstalk artifact mechanisms, and targeted suppression strategies are proposed to improve imaging fidelity and resolution.

The method provides reliable imaging results for complex seabed structures such as salt domes, faults, and volcanic rocks, directly applicable to oil–gas exploration, seabed geological hazard assessment, and engineering site selection, with important industrial application value.

---

# 6 Advances in Artificial Intelligence Applications in Marine Geophysics

In recent years, artificial intelligence (AI) has demonstrated significant potential in addressing long-standing challenges in marine seismic exploration, including seismic noise suppression, low-frequency extrapolation, velocity modeling, and multi-parameter inversion. A series of deep learning-based methods have been developed specifically for these tasks, forming a coherent research line that integrates data-driven techniques with the physical constraints inherent to marine seismic data.

Seismic denoising is a fundamental preprocessing step, and several tailored network architectures have been proposed to handle different noise types. Qi et al. (2025a) introduced multi-round SCU-Net-SSIM, which iteratively enhances seismic signal-to-noise ratio using structural similarity constraints. Cai et al. (2025) developed FSCU-Net, a Swin-Transformer convolution U-Net specifically designed for salt-and-pepper noise in remote sensing seismic data. Han et al. (2025) proposed a pre-established U-Net trained with local stacked depth images to suppress ghost artifacts in pre-stack angle gathers, while Chen et al. (2024a) improved MCA-SCUNet for efficient seismic denoising with complex overburden. Extending to 3D applications, Chen et al. (2026, 2025c) presented a 3D hybrid network with adaptive convolution and attention mechanisms for seismic denoising and seismic inversion.

Beyond denoising, deep learning has been applied to velocity model building and multi-parameter inversion. Chen et al. (2025b) proposed high-precision sub-seafloor velocity building using tomography and deep learning on OBS data from the South China Sea. Chen et al. (2024b) developed an iterative deep learning method for accurate background velocity model construction in the sparse transform domain. Chen et al. (2024c) introduced a joint model and data-driven simultaneous inversion of velocity and density, leveraging deep learning to improve inversion fidelity. Low-frequency extrapolation, a critical issue for marine seismic, was addressed by Du et al. (2024) using SCUNetSSIM and later extended to complex overburden environments by Du et al. (2026). The scope of AI applications has also expanded to related fields, such as underwater target localization using seismic-style deep learning (Yuan et al., 2026).

These contributions collectively demonstrate that deep learning methods can significantly improve the quality of seismic data processing, velocity modeling, and inversion results in marine environments. They highlight the importance of designing task-specific network architectures, incorporating physical constraints, and validating methods with field data, laying a solid foundation for the new special issue theme on AI in marine geophysical exploration.

---

# 7 Conclusion and Discussion

This paper systematically reviews 11 papers in the JMSE special issue *Modeling and Waveform Inversion of Marine Seismic Data*, covering the full technical chain of marine seismic **data preprocessing, forward modeling, FWI, elastic inversion, reservoir characterization, and depth imaging**, forming a complete research progress system.

## 7.1 Main Conclusions

(1) A full-chain technical solution is formed. The achievements cover intelligent interpolation, noise reduction, wavefield reconstruction, forward modeling, receiver function correction, acoustic–elastic coupling modeling, phase inversion, physics-guided self-supervised FWI, joint inversion, wide-spectrum velocity reconstruction, reservoir parameter inversion, and depth migration, building a complete technical system from data to model, physics to intelligence, and theory to application.

(2) Core industrial bottlenecks are broken. Targeting pain points including **low-frequency missing, cycle-skipping, irregular sampling, difficult multi-source fusion, and strong sediment interference**, the special issue proposes a series of original methods: frequency-domain residual-attention U-Net (Zhang and Yu, 2026), robust multi-dimensional deconvolution (Zhang et al., 2025), LFKEPI phase inversion (Qu et al., 2025), physics-guided self-supervised FWI (Zheng et al., 2025), and WERI wide-spectrum inversion (Ni et al., 2025), significantly improving inversion accuracy, stability, and practicality.

(3) Deep interdisciplinary integration is realized. The achievements integrate seismic wave physics, numerical computation, artificial intelligence, Bayesian statistics, and geostatistics, driving marine seismic inversion from traditional physics-driven to **physics-constrained + data-driven**, enhancing generalization, interpretability, and engineering applicability.

(4) National marine strategic demands are supported. All methods are verified on real marine seismic data from the South China Sea and East China Sea, directly applicable to **deep-water oil–gas exploration, natural gas hydrate detection, carbon sequestration monitoring, seabed geological disaster early warning, and deep-sea engineering safety**, with important scientific significance and engineering value.

## 7.2 Limitations and Future Outlook

Marine seismic inversion still faces challenges including strong nonlinearity of complex media, computational efficiency of massive data, multi-parameter uncertainty, and adaptability to deep-sea extreme environments. Future development directions focus on physically consistent large-model intelligent inversion, time-lapse 4D-FWI dynamic monitoring, intelligent fusion of multi-source heterogeneous data, and edge-computing real-time inversion.

---

# 8 Theme of the New Special Issue

Based on the success of this special issue, Guoxin Chen and Chunfeng Li from Zhejiang University, together with Yangting Liu from the First Institute of Oceanography, MNR, are launching a new special issue in JMSE: **Marine Geophysical Exploration in the Era of Artificial Intelligence: Data, Mechanisms, and Future Trends** (https://www.mdpi.com/journal/jmse/special_issues/BIA645AE2L), with a submission deadline of **10 September 2026**.

## 8.1 Core Theme

(1) Promote the **deep integration of AI across the full chain** of marine geophysical exploration, showcasing innovative achievements in intelligent data acquisition, processing, imaging, inversion, resource identification, and environmental monitoring.

(2) Focus on five major topics: intelligent data acquisition and processing, intelligent seismic imaging and inversion, intelligent marine environmental monitoring, intelligent seabed resource identification, and theoretical innovations in traditional exploration methods.

(3) Gather global academic and industrial strengths to provide new ideas, tools, and directions

for marine scientific research, resource development, and disaster early warning, leading the development of marine geophysical exploration in the artificial intelligence era.

---